\documentclass[10pt,letterpaper]{article}
\usepackage[top=0.85in,left=2.75in,footskip=0.75in,marginparwidth=2in]{geometry}

\usepackage[utf8]{inputenc}

\usepackage{cite}

\usepackage{nameref,hyperref}

\usepackage{microtype}
\DisableLigatures[f]{encoding = *, family = * }

\raggedright
\usepackage{changepage}

\usepackage[aboveskip=1pt,labelfont=bf,labelsep=period,singlelinecheck=off]{caption}

\makeatletter
\renewcommand{\@biblabel}[1]{\quad#1.}
\makeatother

\usepackage{lastpage,fancyhdr,graphicx}
\usepackage{epstopdf}
\fancyheadoffset[L]{2.25in}
\fancyfootoffset[L]{2.25in}

\usepackage{color}

\definecolor{Gray}{gray}{.25}

\usepackage{graphicx}

\usepackage{sidecap}

\usepackage{wrapfig}
\usepackage[pscoord]{eso-pic}
\usepackage[fulladjust]{marginnote}
\reversemarginpar

\begin{document}
\vspace*{0.35in}

\begin{flushleft}
{\Large
\textbf\newline{Design and Testing of a Low-Cost 3D-Printed Servo Gimbal for Thrust Vector Control in Model Rockets }
}
\newline
\\
Ekansh Singh
\\
\bigskip
\bf{1} Pope High School, Advanced Math and Science STEM Academy, Marietta, GA, USA 
\\
\bigskip
* ekanshsingh2026@outlook.com

\end{flushleft}

\section*{Abstract}
Thrust vector control (TVC) is a key mechanism for stabilizing rockets during flight, yet conventional implementations remain costly and technically inaccessible to students and hobbyists. This paper presents the design, fabrication, and testing of a low-cost, 3D-printed, servo-driven two-dimensional gimbal developed for model rocket applications. The gimbal underwent more than 60 CAD iterations, with servo selection guided by torque, response time, and stability requirements. A high-speed camera and Fusion 360 parameter simulations were used to emulate dynamic instability, enabling evaluation of angular deflection, servo responsiveness, and structural durability. The results demonstrated stable actuation within ±5°, with response times on the average order of 44.5 ms, while limitations included servo fatigue and pin-joint stress under extended loading. The project highlights the feasibility of student-accessible thrust vector control systems and their potential as a reproducible platform for STEM education and experimental aerospace research.


\section*{Introduction}
\begin{wrapfigure}[17]{l}{57mm}
\includegraphics[width=52.5mm]{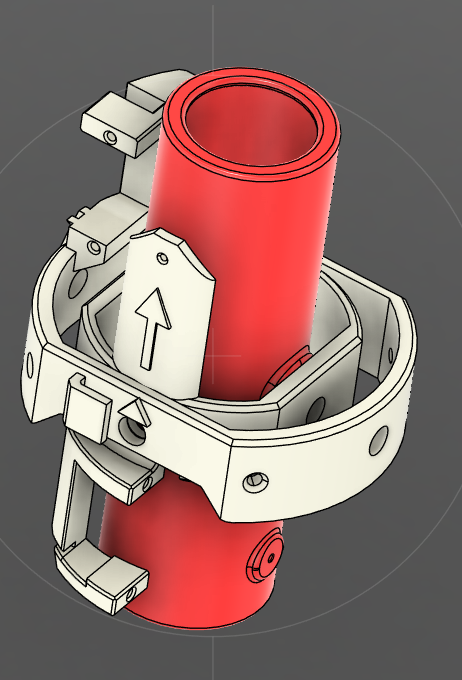}
\captionsetup{labelformat=empty} 
\caption{} 
\label{fig1} 
\end{wrapfigure}

As shown in \textit{(Fig. \ref{fig1}).,} this study focuses on a simplified, 3D-printed gimbal mechanism that demonstrates Thrust Vector Control (TVC) principles at the model rocket scale. 
\marginpar{
\vspace{.7cm} 
\color{Gray} 
\textbf{Figure \ref{fig1}. }Final CAD model of the 3D-printed, two-axis servo gimbal designed for thrust vector control in model rockets. The design incorporates slots for orthogonally mounted micro-servos and a central motor mount, enabling thrust deflection along both pitch and yaw axes.
}
Thrust vector control (TVC) is the process of actively directing the thrust of a rocket engine to maintain stability and control its flight trajectory \cite{Sutton2017}. By altering the direction of exhaust gases relative to the rocket’s center of mass, TVC enables vehicles to counteract aerodynamic disturbances and remain on their intended path. Nearly all modern launch vehicles, from the Saturn V to the Falcon 9, employ some form of TVC to achieve reliable ascent and precise maneuvering in space.
The importance of robust TVC is illustrated by the second demonstration flight of SpaceX’s Falcon 1 rocket in 2007 \cite{Musk2009}. During ascent, the vehicle lost control after its thrust vectoring algorithm failed to adequately stabilize the second stage. The resulting instability led to the premature termination of the mission. This and similar incidents highlight both the technical challenges and critical importance of guidance and control systems in rocketry.
Although rarely studied in formal student research, TVC has gained traction within the hobbyist rocketry community \cite{Tavares2020}. Numerous online demonstrations employ 3D-printed gimbals actuated by micro-servos and controlled with Arduino-based systems, proving that basic thrust vectoring is achievable at a small scale. However, these implementations often suffer from recurring shortcomings. 3D-printed CAD mechanisms are frequently brittle and prone to failure under sustained stress and strain, while inexpensive servos seem to lack the torque and speed necessary for reliable calibration. In addition, many designs are unnecessarily complex, resulting in assemblies that are difficult to fabricate reproducibly and provide limited structural advantage.
This paper presents the design, fabrication, and testing of a reproducible, low-cost, 3D-printed, servo-driven gimbal developed specifically for model rocket applications. By iterating through more than sixty CAD prototypes and testing performance with Fusion 360 simulations and high-speed video analysis, this project demonstrates that meaningful control system research can be achieved at low cost. The goal is to address the fragility, sluggish response, and over-complication often found in hobbyist systems, and to provide a durable, student-accessible platform for hands-on aerospace education and experimental rocketry.

\section*{System Overview}

\begin{wrapfigure}[14]{l}{60mm}
\includegraphics[width=57mm]{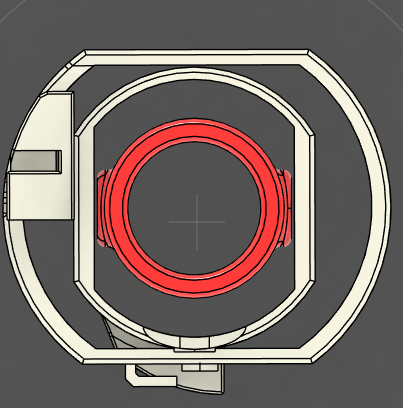}
\captionsetup{labelformat=empty} 
\caption{} 
\label{fig2} 
\end{wrapfigure}

The thrust vector control gimbal consists of a two-axis, servo-actuated mechanism designed to mount beneath a model rocket motor. The system is composed of three primary elements: (1) a central cylindrical motor mount, (2) a lightweight, 3D-printed gimbal frame, and (3) two orthogonally arranged micro-servos that provide independent control over pitch and yaw. Together, these components enable the motor’s exhaust vector to be deflected by up to ±5°, sufficient for stabilizing a small-scale rocket in flight simulations. The servos are arranged at right angles to one another, with one responsible for pitch actuation and the other for yaw. Their motion is transferred directly to the motor mount via reinforced 3D-printed pivot joint pins, minimizing backlash and allowing rapid angular response. The gimbal frame is printed in ABS, chosen for its accessibility and durability, with structural reinforcement incorporated through iterative CAD modifications to reduce deformation and fatigue. This combination of virtual and physical validation ensured both mechanical durability and dynamic responsiveness, crucial factors for successful flight, could be quantitatively and qualitatively assessed. \textit{(Fig. \ref{fig2}).,} presents a top-view schematic of the gimbal, highlighting the orthogonal servo arrangement and central rocket motor mount tube. 
\marginpar{
\vspace{.7cm} 
\color{Gray} 
\textbf{Figure \ref{fig2}. }Bird’s-eye CAD rendering of the two-axis thrust vector control gimbal. 
}
Performance of the system was evaluated using both computational and experimental methods. Stress and strain analyses in Autodesk Fusion 360 were used to identify likely points of failure in the printed frame under simulated thrust loads, while high-speed camera recordings provided frame-by-frame measurements of servo response and deflection angle. This overview serves as a foundation for the subsequent discussion of the design process, testing methods, and performance results. The thrust vector control gimbal functions not only as a stabilization mechanism but also as a scalable platform for aerospace education. By using 3D printing and inexpensive off-the-shelf components, the design lowers the barrier to entry for students and hobbyists who might otherwise be excluded from experimenting with flight control systems. In contrast to commercial, ready-to-fly-kit gimbals, which are costly and inaccessible, this system demonstrates that reliable thrust vectoring can be achieved with a few servos, efficient CAD modeling, open-source analysis tools and programming. As such, the platform enables hands-on learning in areas such as control theory, structural mechanics, and embedded systems, aligning with the broader goal of making aerospace research more approachable in academic and extracurricular contexts. A critical design trade-off lies in the balance between servo torque and angular deflection \cite{Wertz1999}. The 9-gram servos used in the gimbal provide limited torque capacity and precision, insufficient for direct deflection of the motor mount under load. To overcome this, the gimbal employs a leveraged geometric amplification ratio in its pivot arrangement, converting large servo movements into central nozzle deflections of up to ±5°. While this extends the effective control authority of the system, it comes at the expense of increased mechanical stress on the joints and higher demands on structural rigidity. Iterative CAD and simulations were used to optimize this trade-off, ensuring that servos remained within safe operating ranges while the frame maintained durability under amplified loads which rockets would be subject to during flight. The electrical and computerized control system was kept intentionally simple to emphasize accessibility and replicability. Both servos are connected through a prototyping board controlled by a common low-cost micro-controller such as an Arduino Nano. Pulse-width modulation (PWM) signals drive the servos, with independent channels corresponding to orthogonal pitch and yaw actuation. This compact approach allows the gimbal to be integrated into different testing environments, from bench top deflection studies to full rocket flights, without requiring specialized, inaccessible hardware.

\section*{Mechanical Design}
\begin{wrapfigure}[17]{l}{57mm}
\includegraphics[width=60mm]{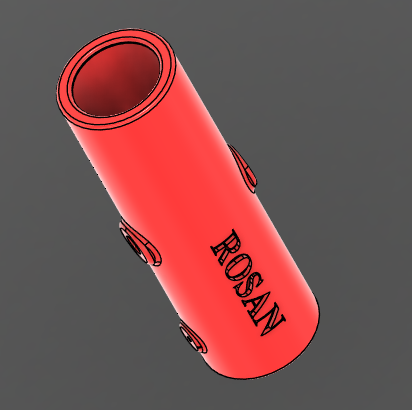}
\captionsetup{labelformat=empty} 
\caption{} 
\label{fig3} 
\end{wrapfigure}

\begin{wrapfigure}[17]{l}{57mm}
\vspace{6.4cm}
\includegraphics[width=60mm]{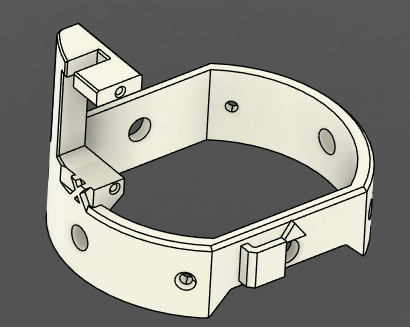}
\captionsetup{labelformat=empty} 
\caption{} 
\label{fig4} 
\end{wrapfigure}

The development of the thrust vector control gimbal was driven by iterative design, with over sixty CAD prototypes produced before arriving at the final configuration. A central challenge throughout the process was reconciling the need for compactness with the structural requirements of a servo-driven gimbal. The final design employs a 35 mm inner cylindrical motor mount, sized to hold a standard 29 mm F-series Estes rocket motor, as shown in \textit{(Fig. \ref{fig3}).,} and a concentrically housed 74 mm outer gimbal frame, \textit{(Fig. \ref{fig4}).,}. This geometry ensures compatibility with most commercial cardboard rocket airframes while leaving adequate clearance for servo placement, pivot joints, and structural reinforcement. Early iterations highlighted the limitations of conventional rectangular gimbal geometries. Designs that used rectangular inner and outer rings, were quickly found to be incompatible with cylindrical rocket airframes, as the corners of the rings interfered with the tube wall and restricted angular deflection along with creating instability. These versions also added unnecessary bulk and mass. By approximately the twentieth part iteration, the design transitioned fully to cylindrical geometry, which simplified integration into the airframe, reduced wasted space, maximized stability, minimized weight, and maintained a relatively simple, replicable design. From that point onward, successive refinements focused on wall thickness, servo clearance, and joint tolerances to maximize durability without compromising manufacturability and low-bulkiness. Material selection played a critical role in these refinements. Initial prototypes were printed in polylactic acid (PLA), which offered ease of printing, extreme accessibility and dimensional stability but proved suboptimal for flight applications in contrast to alternatives. With a density of 1.24–1.25 g/cm³, PLA made the assembly heavier than necessary, and its relatively low resilience raised concerns about light thermal deformation during sustained operation near the motor, in addition to lower structural force tolerance. Subsequent iterations transitioned to acrylonitrile butadiene styrene (ABS), which is less dense 1.04 g/cm³, offers improved impact strength, thermal resistance while staying widely accessible in contrast to materials such as polycarbonate (PC) \cite{Rehnberg2018}. This switch lowered overall structural mass while enhancing durability, enabling the gimbal to endure repeat actuation cycles under thrust and gravitational forces without softening, warping or snapping during flights. Reinforcement ribs and fillets were incorporated into high-stress regions identified through early trial simulations, further improving resistance to bending and fatigue. The pivot pin system also underwent major evolution. Early designs used traditional metal screws to secure the gimbal’s joints, but these introduced excessive friction, added mass, and proved tedious to assemble. Moreover, the concentrated loading of metal-on-plastic interfaces accelerated wear and increased the torque demand on the servos. To overcome these issues, the final system employs 3D-printed snap-in pins with a diameter of 6.35 mm. This geometry provided a balance between stability and smooth rotation, distributing shear stresses across a larger surface area while keeping the mechanism lightweight. The snap-fit design eliminated the need for threading, reduced backlash, and lowered servo loads, ultimately improving responsiveness and repeatability in deflection tests. Together, these geometric, material, and mechanical refinements transformed the gimbal from bulky and impractical early prototypes into a lightweight, robust, and modular system. The final configuration is both structurally efficient and broadly compatible with standard hardware, making it suitable as a test platform for thrust vector control studies in educational and hobbyist contexts.
\marginpar{
\vspace{1cm} 
\color{Gray} 
\textbf{Figure \ref{fig3}. }CAD rendering of the central motor mount. The 35 mm tube, engraved with the ROSAN label, is designed to house a 29 mm Estes F-series motor and serves as the structural core around which the gimbal assembly is built. 
}
\marginpar{
\vspace{3.3cm} 
\color{Gray} 
\textbf{Figure \ref{fig4}. }CAD rendering of the outer-axis gimbal assembly. The design shows the concentric outer frame with orthogonal servo slots and pivot joint holes, enabling ±5° thrust deflection for stabilization in 72 mm airframes.
}
\clearpage 

\section*{Methods and Testing Setup} 

The performance of the thrust vector control gimbal was assessed using both computational simulations and physical experiments. This dual approach enabled characterization of the structural durability of the 3D-printed assembly and the dynamic responsiveness of the servo actuation system. Finite element stress–strain analyses were performed in Autodesk Fusion 360 to replicate thrust loads representative of an Estes F-series motor. A nominal load of approximately 30 newtons applied concentrically to the motor mount \cite{Estes2021}. The simulations produced displacement fields, Von Mises stress distributions, and calculated safety factors under repeated thrust cycles. Stress concentrations consistently appeared at the pivot joints and servo mounting slots, which guided reinforcement strategies such as filleted corners, rib structures, and optimized wall thickness. Physical testing was conducted on a tabletop rig with an Arduino Nano micro-controller generating independent PWM signals for the wire soldered pitch and yaw SG90 micro-servos. The gimbal was actuated through ±5° commands, with deflections recorded using a high-speed camera operating at 120 frames per second (FPS). Frame-by-frame analysis provided measurements of both angular displacement with an accuracy of ±0.5° and servo response times between command input and physical actuation. Failures were defined as a loss of deflection accuracy greater than ±1° or any visible plastic deformation of the assembly. Three metrics were used to evaluate physical performance. The first was deflection range, defined as the maximum angular displacement of the motor mount relative to neutral, mathematically ±5° \cite{Microchip2016}. The second was response time, defined as the elapsed time required for the servo to achieve 95\% of the commanded deflection. Together, ten trials were conducted; these criteria provided a quantitative framework for validating the feasibility of the gimbal and for comparing experimental results against simulation predictions, in addition to a qualitative framework ensuring the assembly stayed intact through physical trials. Both provide both quantitative and qualitative data on TVC frame and servo mount performance under optimal and peak force conditions. 

\begin{wrapfigure}[17]{l}{57mm}
\vspace{1.5cm}
\includegraphics[width=130mm]{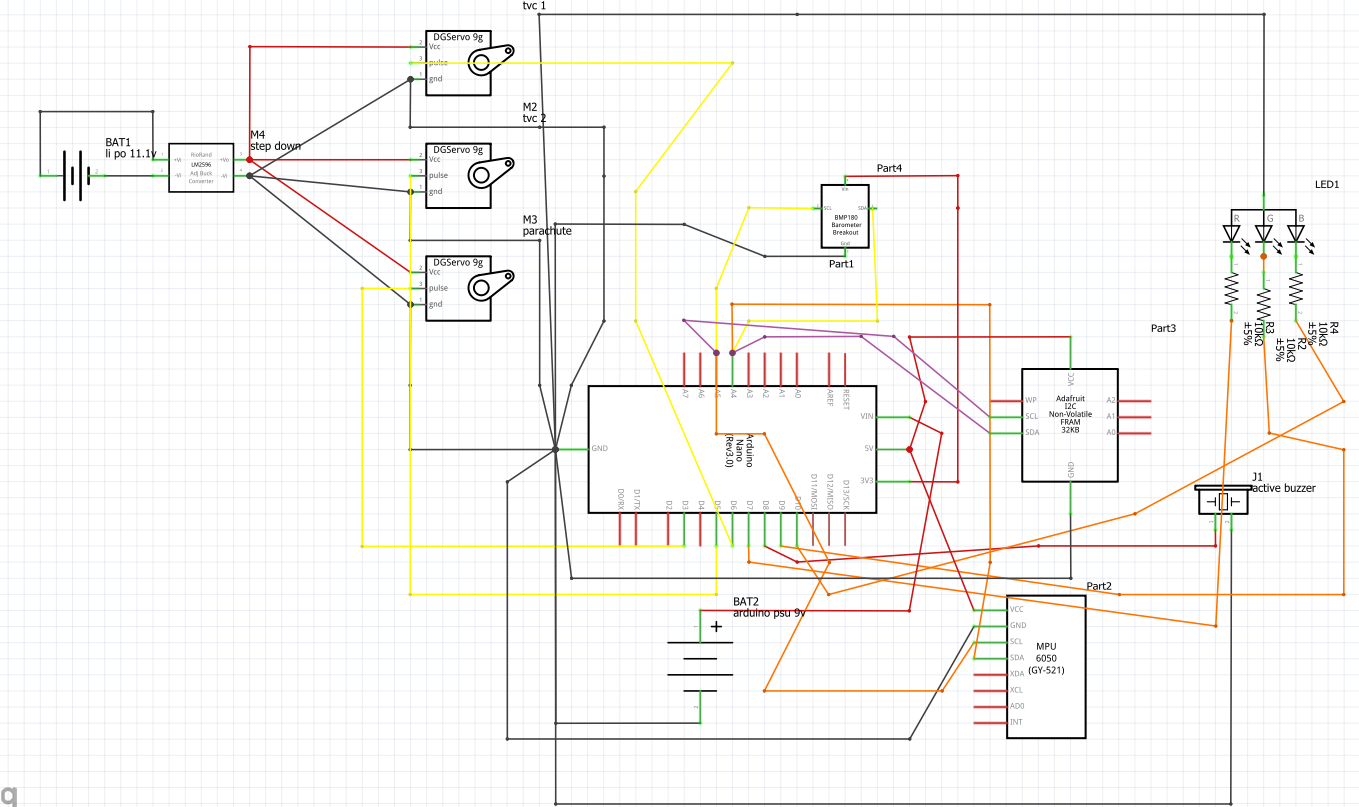}
\captionsetup{labelformat=empty} 
\caption{} 
\label{fig5} 
\end{wrapfigure}

\marginpar{
\vspace{3.3cm} 
\color{Gray} 
\textbf{Figure \ref{fig5}. }Schematic of flight computer assembly. The full computer was not utilized for the tabletop test bed servo response trials; rather, the pitch and yaw servos powered by Li-po cells were.
}
\clearpage 

\section*{Results and Discussion} 
\begin{wrapfigure}[21]{l}{65mm}

\includegraphics[width=63mm]{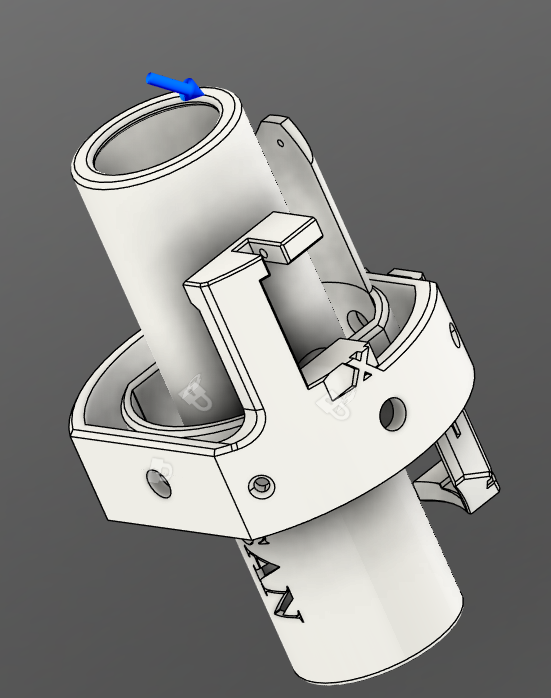}
\captionsetup{labelformat=empty} 
\caption{} 
\label{fig6} 
\end{wrapfigure}

The thrust vector control gimbal was evaluated through a combination of computational simulations and benchtop experimental testing. The results characterize both the structural integrity of the 3D-printed assembly and the responsiveness of the servo-driven actuation system. Static Stress analysis conducted in Autodesk Fusion 360 under a nominal thrust load of 30 N revealed distinct regions of stress concentration near the pivot joints and servo mounting slots, (Fig. \ref{fig6}). Maximum von Mises stress values occurred at the filleted corners adjacent to the motor mount, while the outer frame exhibited comparatively lower stress. These results validated the need for reinforcement ribs and optimized wall thickness in a final design, as previous prototypes often failed.
\marginpar{
\vspace{1cm} 
\color{Gray} 
\textbf{Figure \ref{fig6}. }Static Stress test utilizing concentric force vectors, one displayed, replicating Estes F-15 30 Newton rocket motor. Wider pin slots as compared to prototypes allows for more distributed and higher potential Von Mises stress distributions.
}
Motion was recorded at 120 fps with a high-speed camera, and frame-by-frame analysis yielded both response times and angular accuracy. Across ten trials, the gimbal achieved stable actuation within ±0.2° of the commanded deflection. As shown in (Table \ref{tab:servo_results}) Response times averaged 44.5 ms with a standard deviation of ±2.3 ms. Physical gimbal blemishes were extremely minor, thereby insignificant during one-minute flights. Consistent PWM signals across the ten trials through varying commanded deflections of ±5° attributed to minute angle errors on average, approximately 2 percent. 
Structurally, the ABS frame maintained integrity under thrust loads of 30 N, with stress concentrations localized to predictable regions that were reinforced through iterative design. 

\begin{table}[h]
\centering
\vspace{1cm}
\caption{Servo response times and deflection accuracy for the SG90 micro-servo mounted on the two-axis TVC gimbal (10 trials at $\pm 5^{\circ}$ commands).}
\label{tab:servo_results}
\begin{tabular}{c c c c}
\hline
Trial & Commanded Deflection (°) & Measured Deflection (°) & Response Time (ms) \\
\hline
1 & +5  & +5.1 & 42 \\
2 & -5  & -4.9 & 47 \\
3 & +5  & +5.0 & 44 \\
4 & -5  & -5.1 & 46 \\
5 & +5  & +4.8 & 41 \\
6 & -5  & -5.0 & 45 \\
7 & +5  & +5.0 & 43 \\
8 & -5  & -4.9 & 44 \\
9 & +5  & +5.2 & 48 \\
10 & -5 & -5.0 & 45 \\
\hline
\textbf{Average} & --- & \textbf{$\pm$0.1° error} & \textbf{44.5} \\
\textbf{Std. Dev.} & --- & \textbf{0.12°} & \textbf{2.3} \\
\hline
\end{tabular}
\end{table}

\clearpage

\vspace{1cm}

The combination of mechanical and experimental results demonstrates the feasibility of the proposed gimbal as a low-cost thrust vector control system. Thermal analysis confirmed that ABS provides a modest safety margin but slightly approaches its thermal limit under sustained heating, suggesting that future iterations may benefit from higher-temperature polymers such as polycarbonate (PC). Experimentally, the gimbal achieved accurate ±5° deflections with rapid response times below 50 ms, sufficient for model-scale rocketry and thrust vector stabilization. However, limitations included minor overshoot and undershoot behavior, gradual wear of pivot pins under repeated cycling, and high Von Mises stress regions surrounding pivot pins under load. These findings emphasize both the promise and the constraints of using inexpensive SG90 micro-servos and 3D-printed components in aerospace control applications. Overall, the results highlight the viability of a student-accessible, reproducible thrust vector control platform, offering both functional performance for hobbyist rocketry and educational value in demonstrating core aerospace engineering principles in late high school engineering. 

\section*{Conclusion}
This work presented the design, fabrication, and evaluation of a low-cost, 3D-printed thrust vector control gimbal for model rocket applications. More than sixty CAD iterations and material refinements led to a compact, durable design capable of deflecting a rocket motor by ±5°. Static stress simulations under a 30 N thrust load confirmed that the ABS frame provided sufficient strength and stress distribution for model rocket flights, while thermal analysis indicated only a modest safety margin below the material’s glass transition temperature. Experimental trials using SG90 micro-servos demonstrated accurate actuation within ±0.2° and rapid response times averaging 44.5 ms, consistent with expected servo performance under load. The results highlight that reliable thrust vector control can be achieved with inexpensive hardware and 3D-printed manufacturing, providing both functional stability for model rockets and a reproducible platform for STEM aerospace education. Limitations of the current system include possible servo fatigue during flight, localized stress concentrations leading to long-term strain at pivot pins, and the degradability of plastic pivot pins. Future work will focus on broader TVC rocket control through closed-loop control integration, pyrotechnic air-brakes, and further validation through powered rocket flights.

\end{document}